\documentclass[double]{article}
\usepackage{times}
\usepackage{algorithm}
\usepackage{algorithmic}
\usepackage{wrapfig}
\usepackage{verbatim}
\usepackage{amsmath}
\usepackage{graphicx}
\usepackage{subcaption}
\usepackage[active]{srcltx}
\usepackage{color}
\usepackage{lineno}
\usepackage{dirtytalk}
\usepackage{amsthm}
\usepackage{authblk}
\usepackage{listings}
\usepackage{tikz}
\usetikzlibrary{shapes.geometric, arrows}

\usepackage{epsfig,graphicx,amsfonts}
\usepackage{amsmath,amsthm,amssymb}
\usepackage{xcolor}

\usepackage{adjustbox}
\usepackage{multirow}
\usepackage{setspace}
\usepackage{hyperref}
\usepackage{comment}

\long\def\comment #1\commentend{}

\begin{document}
%\documentstyle[amsfonts]{article}

% Following creates the title page
\title{\Large The Academic Midas Touch: An Indicator of Academic Excellence}
\author{Ariel Rosenfled$^{1}$, Ariel Alexi$^{1}$, Liel Mushiev$^{2}$, and Teddy Lazebnik$^{3*}$\\ \(^1\) Department of Information Science, Bar Ilan University, Ramat Gan, Israel\\ \(^2\) Department of Computer Science, Holon Institute of Technology, Holon, Israel\\ \(^3\) Department of Cancer Biology, Cancer Institute, University College London, London, UK\\ \(*\) Corresponding author: lazebnik.teddy@gmail.com}
%\author{Anonymous for review}

\date{ }

\maketitle 

\begin{abstract}
\noindent
The recognition of academic excellence is fundamental to the scientific and academic endeavor. However, the term \say{academic excellence} is often interpreted in different ways, typically, using popular scientometrics such as the H-index, i10-index, and citation counts. In this work, we study an under-explored aspect of academic excellence -- researchers' propensity to produce highly cited publications. We formulate this novel perspective using a simple yet effective indicator termed the \say{Academic Midas Touch} (AMT). We empirically show that this perspective does not fully align with popular scientometrics and favorably compares to them in distinguishing award-winning scientists.  \\ \\ 

\end{abstract}

\maketitle \thispagestyle{empty}

% Begin using page numbers and a header
\pagestyle{myheadings}
% \markboth{Draft:\today}{Draft:\today}
\setcounter{page}{1}% reset page number to 1

\section{Introduction}
\label{sec:introduction}

%Excellence is important
Evaluating and assessing researchers' academic performance is crucial to maintaining the quality and integrity of research, promoting scientific progress, allocating resources effectively, and ensuring accountability in the research community \cite{researchers_evaluation, lindahl2023conscientiousness, predeict_reseacher_influence, ms_dropping, rotem2021open, sahudin2023determinants}. A central issue within this realm is the identification and recognition of academic excellence \cite{academic_excellence_1,paper_impact}. Specifically, it is important to acknowledge researchers who perform outstanding science. This, in turn, can reinforce high-quality scientific progress and catalyze the pursuit of academic distinction, visibility, and impact.

%Citations are key for identification
Due to its amorphous nature, the term \say{academic excellence} is often interpreted in various ways which need not necessarily align. Indeed, over a hundred different researcher-level performance indicators, which are typically at the basis of academic excellence identification and exploration, have been proposed and evaluated in prior literature \cite{wildgaard2014review, sziklai2021ranking, salmi2011road, kulczycki2017toward}. 
For example, Rodriguez et al. \cite{rodriguez2011measuring} showed that the number of publications, citations, and top one percent most cited publications correlate with Nobel Prize achievements. Similarly, \cite{robinson2019relative} show that H-index and i10-index are useful indicators of medical consultants' success in the United Kingdom and \cite{kpolovie2017research} showed these two indexes are also useful to distinguish excellent academic departments and institutes.
Amongst these measures of excellence, citation-based scientometrics seems to be the most widely accepted quantitative measure for assessing scientific excellence \cite{borchardt2018academic,alexi_master, massucci2019measuring}.
These measures range from simple ones such as the number of publications at the top 5\% most frequently cited publications in the field \cite{frequently_cited_publications}, the number of publications in highly-cited journals \cite{publications_in_highly_cited_journals}, the number of publications which received at least 10 citations (aka i10-index) \cite{i_10_index_1, i_10_index_2}, a factoring of both citations and the number of publications such as the various versions of the h-index \cite{koltun2021_h_index,h_idx_2, h_index3}, to more sophisticated ones such as the g-index \cite{discussion_1}, the scientist impact factor  \cite{discussion_2}, and the u-index \cite{discussion_4}, to name a few. Unfortunately, as most researchers tend to agree, defining \say{excellence} in a standardized and consistent way presents serious difficulties
\cite{academic_excellence_2}. Specifically, each researcher-level indicator reflects just one particular dimension of the general concept of research performance or excellence \cite{citation_based_indicator_1, citation_based_indicator_2}. Consequently, the use of only a single indicator to gauge the overall academic performance of a researcher may provide an incomplete picture, and thus a combination of the various types of indicators is needed in order to offer policymakers and evaluators valid and useful assessments \cite{Researchers_Performance_Indicators_1, Researchers_Performance_Indicators_2}.

%consistency
An excellent researcher is not simply one whose work scores highly on the above metrics \cite{vinkler2010evaluation}. Prior literature has suggested multiple aspects deemed desirable, characteristic, and perhaps defining, of excellent researchers such as student supervision \cite{serenko2022scientometric}, funding acquisition \cite{glanzel1994little}, the ability to generate disruptive science \cite{leibel2024we}, etc. In this work, we study an under-explored aspect of academic excellence -- \textit{researchers' propensity to produce highly cited publications}. Specifically, we argue that an excellent researcher is expected to present sustained excellence characterized by a higher propensity to produce highly cited publications compared to peers. We formalize this notion using a simple indicator measuring the portion of highly-cited publications within a researcher's body of work. We term this indicator the \say{Academic Midas Touch} (AMT), drawing inspiration from the famous tale of King Midas from Greek mythology. Specifically, a researcher who produces only highly-cited publications can be considered to have the \say{Midas touch} -- akin to how King Midas turned everything he touched to gold. 
% , -- and adapt it to the researcher evaluation context. Intuitively, an AMT refers to a
To the best of our knowledge, a researcher's tendency to produce highly-cited publications (i.e., \say{golden} publications), is an under-explored perspective on the relationship between productivity and impact, which need not necessarily be well represented within the existing evaluation frameworks.   
Following the formal introduction of the AMT indicator provided next, we present a thorough empirical investigation of both award-winning and non-award-winning mathematicians ($N=8,468$).  

The article is organized as follows: Section \ref{sec:formal} formally defines the AMT indicator. Then, in Section \ref{sec:methods}, we present our investigation of the field of Mathematics. Finally, Section \ref{sec:discussion} interprets our results in context, discusses the main limitations of our work, and highlights potential future work directions.

\section{The Academic Midas Touch}\label{sec:formal}

% Since the definition of a publication's impact is, by itself, a point of contentious \cite{publications_impact}, 

% We use the function $\mathcal{G}(\cdot)$ to measure the \say{quality} associated with an individual publication and calculate its average across a researcher's body of work.
% Formally, let \(s\) be a researcher and let \(p\in P\) denote a publication within her body of work. A researcher's AMT is defined as follows:

% \begin{equation}
% C(s) := \frac{1}{|P|}\sum_{p \in P} \mathcal{G}(p).
% \end{equation}\label{eq:prime}

% Note that $\mathcal{G(\cdot)}$ may be defined in many different ways depending on the specific characteristics one wishes to consider when determining the \say{quality} of a given publication. For example, it may consider the scientometrics associated with the publication outlet \cite{publication_outlet}, the number and positioning of authors \cite{positioning_of_authors}, altmetrics \cite{altmetrics_1, altmetrics_2}, etc. 

Formally, we represent each publication within \(s\)'s body of work (\(p\)) as a series $p:=c_0, c_1, \ldots$ where $c_i$ is the number of citations $p$ has accumulated in the first $i$ years since its publication.
A publication is deemed \say{highly-cited} (i.e., $\mathcal{G}(p)=1$) if and only if it has accumulated at least $y$ citations over the first $x$ years since its publication, and $\mathcal{G}(p)=0$ otherwise.

Based on the above, each researcher is then represented using the portion of publications within his/her total body of work that qualify as highly-cited publications using the following equation - 

\begin{equation}
AMT(s) := \frac{1}{|P|}\sum_{p \in P} \begin{cases}
			1, & c_x \geq y \\
0, & \text{otherwise}
		 \end{cases}
\end{equation}\label{eq:AMT}
where $x$ and $y$ are configurable hyper-parameters denoting the \say{time threshold} and \say{citations threshold}, respectively. For an empty set of publications (i.e., \(|P| = 0\)), we define \(AMT(s) := 0\). 

% Simply put, this measure captures the portion of a researcher's total papers that reach a certain citation threshold at a given time frame, when adopting the definition of Eq. (\ref{eq:AMT}).

\section{Investigation}
\label{sec:methods}

\begin{figure}
    \centering
    \includegraphics[width=0.99\textwidth]{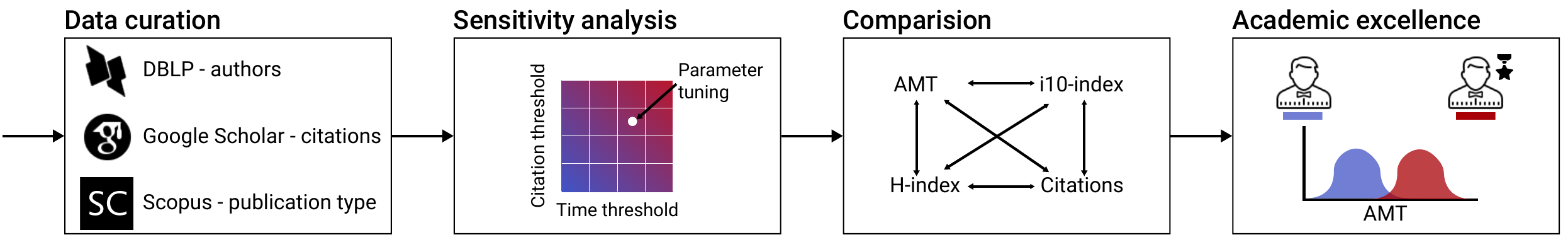}
    \caption{A schematic view of the empirical evaluation process.}
    \label{fig:flow}
\end{figure}

We conducted a four-phased empirical evaluation of mathematicians' propensity to produce highly-cited works, as reflected by AMT, using a sample of 8,468 mathematicians. 
First, we discuss the data curation and processing procedures along with descriptive statistics of the sample. Second, we explore the parameter sensitivity of our definition for highly-cited publications and perform parameter tuning to determine appropriate parameters for the field of Mathematics. Third, we contrast the observed propensity to produce highly-cited publications with popular scientometrics (H-index, i10-index, and citation count) through pair-wise correlation testing. Finally, we examine the propensity to produce highly-cited and its relation with researchers' gender, affiliation continent, academic age and award-winning status. Finally, we examine for possible differences between award-winning mathematicians and their age and productivity matched peers in terms of popular scientometrics (H-index, i10-index, and citation count). 
% Specifically, we show that AMT can serve as a useful indicator for statistically distinguishing between these two groups, whereas other popular scientometrics (i.e., H-index, i10-index, and citation count) do not. 
Fig. \ref{fig:flow} presents a schematic view of the empirical evaluation process.

\subsection{Data curation and processing}
\label{sub_sec:data}

For our empirical investigation, we used three popular academic datasets: DBLP\footnote{\url{https://dblp.org/}}, Google Scholar\footnote{\url{https://scholar.google.com/}}, and Scopus\footnote{\url{https://www.scopus.com/home.uri}}. 
First, all researchers indexed in DBLP were extracted as of June 2023. Note that DBLP is a bibliometric dataset that focuses on the computational domain (i.e., Mathematics, Computer Science, and Engineering) \cite{dblp_good_1,dblp_good_2,dblp_good_3,dblp_good_4} and is considered by many as the state of the art in coverage and accuracy \cite{rosenfeld2023dblp}. In order to single out researchers who primarily publish in the field of Mathematics, i.e., mathematicians, we rely on Scopus's journal subject classification system \cite{classification_system_1, classification_system_2}. Specifically, a publication is considered to be \say{mathematical} if the journal it was published in was classified as mathematical one according to the Web of Science (WOS) index \cite{wos_idx}. For our sample, we consider researchers who published at least five journal articles, of which more than 50\% are classified as mathematical over a time span of at least five years. Overall, 8468 mathematicians were selected for consideration.  
For each of these mathematicians, we extracted their publication records using Google Scholar along with the automatically calculated H-index, i10-index, and citation count. Google Scholar is considered to have the best publication coverage compared to other indexes \cite{gs_3}, and thus, it was chosen for this purpose.
% There are multiple scientometrics for measuring the academic impact of a researcher's body of work with the most commonly used one being the h-index, i10-index, and citation count \cite{pop_metrices_2,pop_metrices_1,ciatation_indices_1,ciatation_indices_2,ciatation_indices_4}. These scientometrics are available as part of a researcher's Google Scholar profile and were extracted as part of our data curation phase.

We further identified each researcher's main affiliation (based on each Google Scholar's profile) and classified it to Europe, North America, Asia, Africa, Oceania, or Other/Unknown. Overall, 3988 mathematicians (47.1\%) are affiliated with European-based universities, 2439 (28.8\%) are affiliated with North American-based universities, 1118 (13.2\%) are affiliated with Asian-based universities, 398 (4.7\%) are affiliated with African-based universities, 212 (2.5\%) are affiliated with Oceania-based universities, and 313 (3.7\%) are affiliated with other or unknown universities. 
Furthermore, we use the gender identification model proposed by \cite{gender_model}, which was trained on around 100 million pairs of names and gender association, and a confidence threshold of 95\% to assign each researcher with an estimated gender. Our sample consists of 7460 (88.1\%) male, 899 (10.6\%) female, and 110 (1.3\%) unknown mathematicians. 
The academic age (i.e., years since first publication) of the mathematicians in our sample ranges from 3 to 38 years with a mean and standard deviation of \(24.18 \pm 4.53\). On average, mathematicians in our sample published \(52.47 \pm 23.09\) publications with an average of \(2.24 \pm 2.91\) publications each year. Overall, 486,622 publications were considered.

\subsection{Sensitivity and Parameter Tuning}

\begin{figure}
\centering
\includegraphics[width=0.7\textwidth]{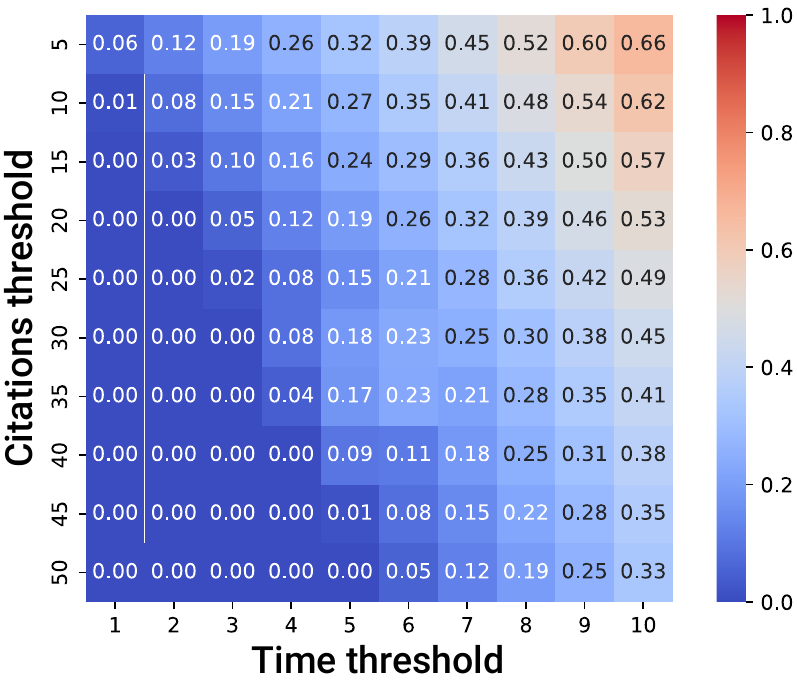}
\caption{The average AMT score in our sample as a function of \(x\) (time threshold)  and \(y\) (citations threshold).}
\label{fig:heatmap}
\end{figure}

Eq. (\ref{eq:AMT}) consists of two hyper-parameters -- \(x\) and \(y\). In order to understand the sensitivity of the formula and identify a sensible tuning for the hyper-parameters, we explore several settings and report the average AMT score across the mathematicians of our sample. Namely, for each combination of \(x\) and \(y\) values, we computed the average AMT score over the 8,468 mathematicians population. As can be observed in Fig. \ref{fig:heatmap}, the higher the values set for $x$ and $y$ -- the higher the average AMT score in the sample. This result is very natural as the number of citations accumulated to a publication is monotonically non-decreasing over time. Using a least mean square approach \cite{lma}, we fit the results with a linear function obtaining: \(C = 0.0438 + 0.0659x - 0.0096y\) with a solid coefficient of determination of \(R^2 = 0.837\). That is, as the time threshold \((x)\) increases and as the citation threshold \((y)\) decreases, the average AMT score increases in an (almost) linear fashion. 

For our subsequent evaluation, we use \(x = 3\) and \(y = 15\). These values were chosen for the following reasons: First, it is often argued that 10 citations are considered to be a worthy threshold to indicate that a publication has been accepted by the academic community \cite{citation_bigger_10} as also mirrored by the popular i10-index \cite{i_10_index_1, i_10_index_2}. As such, we chose to set a higher standard (by 50\%) for what will be considered to be a highly-cited publication. Second, three years are, arguably, enough exposure time to allow a publication to be cited in other researchers' subsequent publications \cite{citation_time}. Third, using these parameters, on average, 10\% of a researcher's publications are considered highly-cited, providing a reasonable balance between ordinary and extra-ordinary (i.e., highly-cited) publications.  Finally, using these parameters, the AMT score distribution in our data does not seem to statistically differ from a Normal distribution using the Shapiro–Wilk Normality test \cite{normal_test}, at $p = 0.188$. 

\subsection{Statistical analysis}

Considering academic age, we computed the Pearson correlation \cite{pearson} between researchers' academic age and their AMT scores, showing a weak positive, yet statistically significant, correlation of 0.17 at \(p = 0.008 \). A statistically significant gender-based difference in AMT scores was also recorded using a Mann-Whitney U test \cite{utest}, demonstrating a higher male tendency to produce highly-cited publications -- an average AMT score of \(0.39 \pm 0.04\) vs \(0.38 \pm 0.03\) with \(p = 0.047 \). Lastly, for the researchers' affiliation, we used the Kruskal–Wallis test with Bonferroni post-hoc correction \cite{kw_test} and found the Europe and North America are associated with statistically significantly higher AMT scores compared to Asia and Africa with all \(p\) values below 0.05. Europe and North America do not statistically differ.

Table \ref{table:corrolations} reports the pair-wise Pearson correlations between AMT, H-index, i10-index, and citation count considering each researcher's complete body of work. As can be observed from the Table, despite the high correlation between the H-index, i10-index, and total number of citations (correlations range between 0.62 and 0.81), the AMT scores seem to only moderately correlate with other examined metrics (correlations range between 0.34 and 0.58). Presumably, this result is an indication that the propensity to publish highly-cited publications captures a slightly different notion of academic performance which is similar, yet not entirely aligned, with popular metrics.

\begin{table}[!ht]
\centering
\begin{tabular}{c|cccc}
 & \textbf{C} & \textbf{H-index} & \textbf{i10-index} & \textbf{Citations} \\ \hline
\textbf{C} & 1 & 0.44 (\(<0.001\)) & 0.34 (\(<0.001\)) & 0.58 (\(<0.01\)) \\
\textbf{H-index} &  0.44 (\(<0.001\)) & 1 & 0.81 (\(<0.01\))&  0.62 (\(<0.001\)) \\
\textbf{i10-index} & 0.34 (\(<0.001\)) & 0.81 (\(<0.01\)) & 1 & 0.75 (\(< 0.001\))  \\
\textbf{Citations} & 0.58 (\(<0.01\))  & 0.62 (\(<0.001\)) &0.75 (\(< 0.001\))  & 1  
\end{tabular}
\caption{Pair-wise Pearson correlations between the AMT and popular scientometrics. The results are shown in brackets as the correlation coefficient and the p-value.}
\label{table:corrolations}
\end{table}

We further consider a sample of 100 award-winning mathematicians who received at least one of the following distinctions: Fields Medal\footnote{According to the international mathematical union, the Fields Medal is awarded every four years to recognize outstanding mathematical achievement for existing work and for the promise of future achievement. It is considered the most prestigious award one can obtain for a mathematical work.}, Abel Prize\footnote{The Abel Prize recognizes pioneering scientific achievements in mathematics, established by the Norwegian Parliament (The Storting) in 2002.}, or Wolf Prize\footnote{The Wolf Prize has been provided since 1978 to outstanding scientists and artists from around the world by the Wolf Foundation.}. The mathematicians selected for this analysis, which we refer to as the award-winning sample, were chosen at random subject to their having a Google Scholar profile. % A full list is provided in the Appendix.

\begin{figure}[!ht]
    \centering
    \begin{subfigure}{.49\textwidth}
        \includegraphics[width=0.99\textwidth]{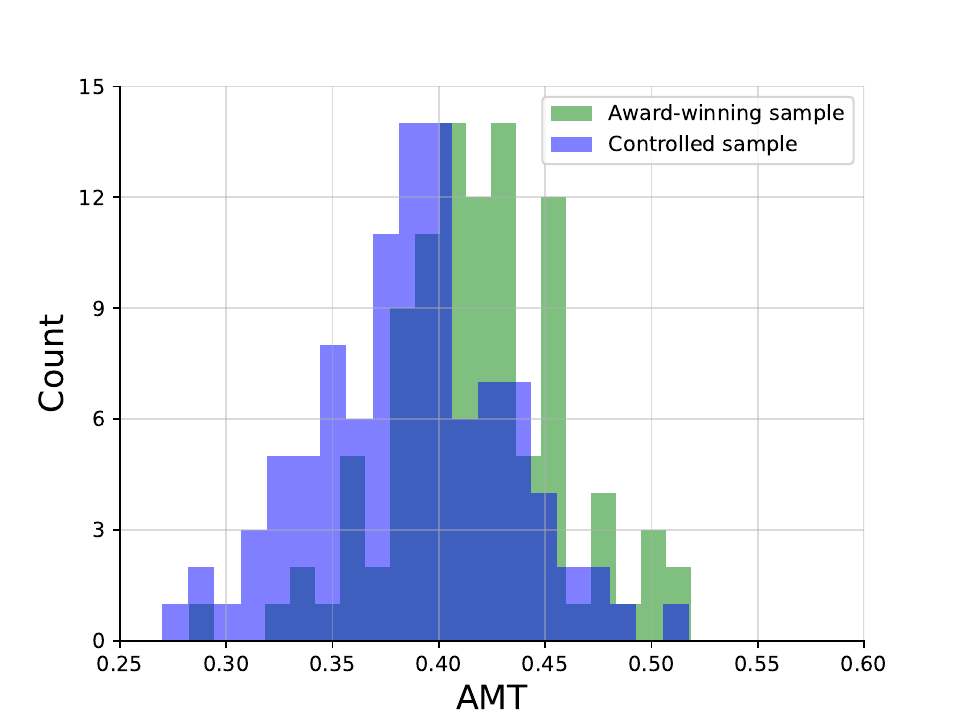}
        \caption{Pre-prize-winning.}
        \label{fig:compare_pre_prize}
    \end{subfigure}
    \begin{subfigure}{.49\textwidth}
        \includegraphics[width=0.99\textwidth]{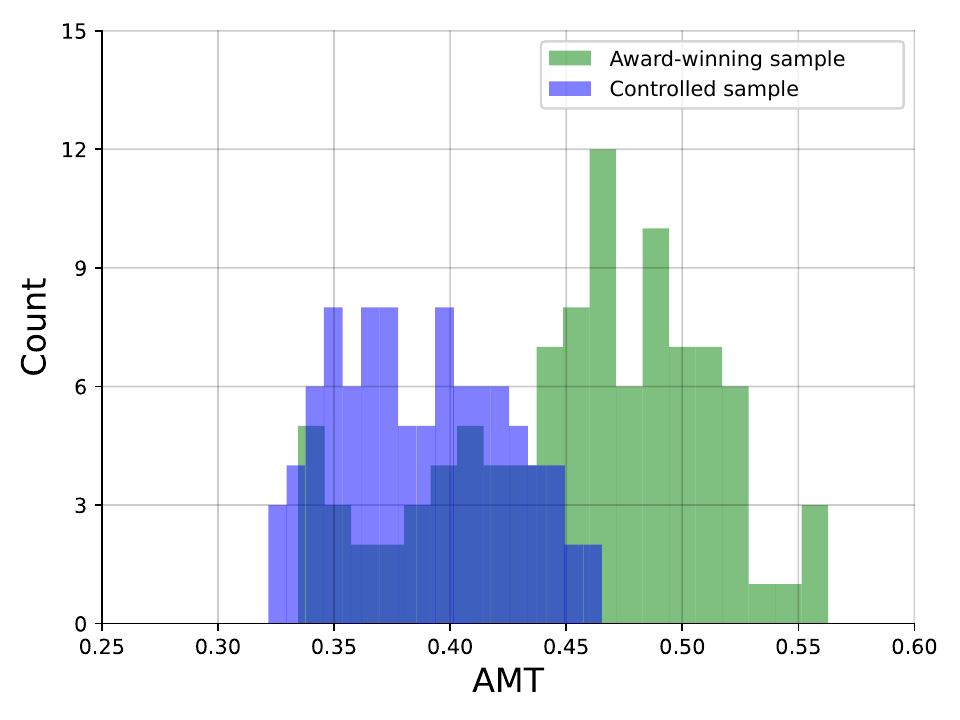}
        \caption{Post-prize-winning.}
        \label{fig:compare_post_prize}
    \end{subfigure}
\caption{The AMT score distributions of the award-winning sample (in green) and the control group (in blue). }
\label{fig:compare}
\end{figure}

In order to examine whether award-winning mathematicians present a distinguished propensity to produce highly-cited works, we examine the AMT distribution of the award-winning sample and compare it to an age and productivity-matched group of non-award-winning mathematicians. The distributions are considered at two time points: the year of the award provision, and the last time point in the data. Specifically, we devise a matched group of mathematicians \(n=100\) who do not statistically differ from the award-winning group, using paired testing, in terms of academic age and number of publications at \(p = 0.328\). Fig. \ref{fig:compare} depicts the AMT score distributions of the award-winning sample (in green) and the matched group (in blue). As can be observed from the figure, the award-winning group demonstrates a very different AMT score distribution, both at the time of award provision and as of the final time point in the data, where the award-winners are centered around higher AMT scores than the control group. Indeed, the two groups are statistically different, with the award-winning mathematicians demonstrating higher average AMT scores both at the prize provision year (\(0.42 \pm 0.04\) vs \(0.39 \pm 0.04\)) and the final year in the data (\(0.45 \pm 0.06\) vs \(0.39 \pm 0.05\), at \(p = 0.014\) and \(p = 0.042\)). A similar statistical comparison between the groups based on popular scientometrics (H-index, i10-index, and citation count), does not point to any statistically significant differences between the two groups at the two time points examines -- the prize year (\(p = 0.079, 0.288, 0.147\)) and the last year in the data (\(p = 0.069\), \(p= 0.115\), and \(p=0.163\), respectively). The results are summarized in Table \ref{table:differances}. 

\begin{table}[!ht]
\centering
\begin{tabular}{llcccc} \hline \hline
\textbf{Time} & \textbf{Scientometric} & AMT & H-index & i10-index & Citation count \\ \hline \hline
Pre-prize year & Relative difference & \textbf{7.59\%} & 6.13\%  & 2.72\%  & 5.69\% \\
Pre-prize year & $p$-value & \textbf{0.01}4 & 0.079 & 0.288 & 0.147 \\ \hline
2023 & Relative difference & \textbf{13.33\%} & 11.07\%  & 4.72\%  & 6.83\% \\
2023 & $p$-value & \textbf{0.042} & 0.069 & 0.115 & 0.163 \\ \hline \hline
\end{tabular}%
\caption{The relative difference (first row) and the statistical significance of the difference (second row) between the mean of the award-winning sample and the control group using the AMT and popular scientometrics (columns). }
\label{table:differances}
\end{table}

\section{Discussion and Conclusion}
\label{sec:discussion}

In this study, we formalized a novel perspective of academic excellence that captures the expectation from prominent researchers to produce highly-cited publications at a higher rate than others. We termed the resulting indicator -- the Academic Midas Touch (AMT). In our simplistic yet effective instantiation of this intuition, we measure the rate of publications one has made that have attracted considerable academic attention (i.e., at least 15 citations) over a short period since their publication (i.e., three years), as indicated by two tunable hyper-parameters. Similar to popular scientometrics such as the H-index, AMT explicitly considers a particular aspect in the relationship between productivity and impact. However, as shown in our empirical evaluation, AMT seems to capture a slightly different viewpoint than popular scientometrics, and brings about favorable indicative properties. 
In particular, using extensive data from the field of Mathematics, we show that AMT scores correlate, but do not fully align, with the H-index, i10-index, and citation counts while favorably comparing to them in distinguishing highly acclaimed, award-winning, mathematicians from others. Taken jointly, these results seem to suggest that the propensity to produce highly-cited publications is a valid and arguably valuable perspective for the distinction of academic excellence that can complement established scientometrics.  

It is important to note that this work has several limitations that offer fruitful opportunities for future work. First, our empirical evaluation focuses on a sample of mathematicians identified through their journal publications. Since the delineation of any scientific field is unclear and journals’ boundaries need not necessarily align with those of any given field of study, journal subject classifications may not align with one's expectations \cite{aviv2023logical}. In other words, various other definitions of which scientists should be considered as \say{mathematicians} could be applied, leading to potentially different results. Second, as different scientific fields may have irregular publication patterns, citation practices, and evaluation criteria, our results may not generalize well outside the field of Mathematics.  As such, the exploration of AMT in additional scientific fields seems merited. Third, our mathematical definition of AMT takes a simple functional. Alternative formulations should be considered in the future. Last, it is important to note that, similar to other researcher-level performance indicators, AMT does not fully capture the multidimensional nature of academic conduct such as collaborations, mentoring, and societal impact which are, by themselves, highly complex and multifaceted. As such, it is intended as a complementary indicator which is to be considered in tandem with others. 
 
\bibliography{biblio}
\bibliographystyle{apalike}

\end{document}